# Impacts of The Radiation Environment At L2 On Bolometers Onboard The Herschel Space Observatory

B. Horeau[1], O. Boulade[1], A. Claret[1], H. Feuchtgruber[2], K. Okumura[1],
P. Panuzzo[1], A. Papageorgiou[3], V. Revéret[1], L. Rodriguez[1], M. Sauvage[1]

*Abstract*—We present the effects of cosmic rays on the detectors onboard the Herschel satellite. We describe in particular the glitches observed on the two types of cryogenic far-infrared bolometer inside the two instruments PACS and SPIRE. The glitch rates are also reported since the launch together with the SREM radiation monitors aboard Herschel and Planck spacecrafts. Both have been injected around the Lagrangian point L2 on May 2009. This allows probing the radiation environment around this orbit. The impacts on the observation are finally summarized.

## I. Introduction

SEVERAL future spatial missions are planned to be placed at the Sun-Earth Lagrangian point L2 due to its interesting properties (thermal properties and low radiative level). In this context the launch of the Herschel Space Observatory by the European Spatial Agency is an opportunity to probe the radiation environment at L2. The radiation monitors aboard Herschel and Planck satellites, both injected at this orbit, allow to study the space environment and monitor the impacts on detectors aboard Herschel Space Observatory.

Since the effect of the high-energy particles may impact on the scientific observations, the identification of theses effects becomes determinant for reaching the scientific goals of the mission. Thus ground radiation tests are often a key issue in the instrument development (e.g. [1] and [2]). In this paper we describe the glitches detected on the two types of far-infrared bolometers inside PACS and SPIRE instruments in orbit. We also describe the radiation environment around L2 thanks to the Standard Radiation Environment Monitors (SREM) since the launch in May 2009. We then report the count rates of the cosmic rays observed in PACS and SPIRE photometer based on bolometers and in PACS spectrometer based on Ge:Ga photoconductors together with the most relevant SREM channels.

We structured our work as follows: in Section 2 we briefly describe the Herschel Space Observatory; in Section 3 we sum up the space environment at L2. We further compare the SREM channels data aboard Herschel and Planck spacecraft; in Sections 4 and 5 the glitches observed on the bolometers are described and analysed, for the PACS and SPIRE photometer respectively; in Section 6 we report the glitch rate on the PACS instruments and the SPIRE photometer; in Section 7 we correlate the glitch rate with the SREM data. Finally, in Section 8, we sum up the impacts of the cosmics rays on the observation time.

## II. The Herschel Space Observatory

### A. The Mission

The Herschel Space Observatory is the fourth of the original cornerstone missions in the ESA Horizon 2000 science plan. It was launched on May 14, 2009 by an Ariane 5 rocket with the Planck spacecraft. Both satellites were injected in a Lissajous orbit around the L2 point. Herschel was designed for an operation time of 3.5 years. The science payload comprises three instruments: two direct detection cameras/medium resolution spectrometers, PACS and SPIRE, and a very high-resolution heterodyne spectrometer, HIFI, whose focal plane units are housed inside a superfluid helium cryostat. The instruments are devoted to spectroscopic and imaging observations in the 60μm to 670μm wavelength range. Stars in the early phases of formation in molecular clouds in the galaxy and star-forming galaxies at high redshifts (up to z~5), covering the epochs of the bulk of the star formation in the universe, emit most of their energy in the Herschel spectral range. The prime science objectives are study of the interstellar medium and the star formation history of the universe, and of the galaxy evolution and the cosmology during the last 10Gyr. For more detailed descriptions, see [3] for the Herschel mission, [4] for the PACS instrument, and [5] for the SPIRE instrument.

### B. The Instruments

This work uses the data extracted from the PACS instrument, both the spectrometer and the photometer, and the SPIRE photometer only.

The PACS instrument employs for the spectrometer two

Ge:Ga photoconductors (stressed and unstressed) with 16*25 pixels, each, and two filled silicon bolometer's focal planes with 16*32 (the red camera) and 32*64 pixels (the blue camera), respectively, for the photometer. They perform integral-field spectroscopy and imaging photometry for the short (the blue channel from 60 to 125 µm) and the long (the red channel from 130 to 210 µm) wavelength regimes [4].

The cameras of the photometer are made of a mosaic of 3-side buttable bolometer arrays (16x16 pixels each and 750µm pixel pitch, see Fig. 1). The detection principle in the PACS bolometer arrays is the resonant absorption of the submillimeter electromagnetic radiation with $\lambda/4$ cavities [6]. They use two silicon chips. The first contains the absorbing insulated meshes (the pixels) with thermometers and the second contains the reflector (in gold), the cold CMOS readout electronic and the multiplexing circuit. The performance of the focal plane is detailed in [7].

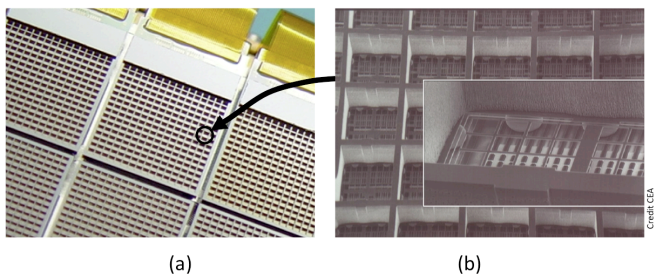

Fig. 1. (Color online) (a) View of the PACS 16*16 bolometer arrays. (b) Microscope view of a single bolometer inside an array. The grid is linked to the inter-pixel walls via narrow silicon beams (2µm wide). We can see them in the zoom-in. We can also recognize the indium bumps under the inter-pixel wall. They define the cavity height (20µm thick for PACS) and make the thermal link between the silicon detection layer and the substrate. They conduct also the electrical signals between the thermometer and the cold-stage electronics.

The SPIRE photometer uses three arrays of feedhorn-coupled NTD-germanium bolometers in order to carry out broadband photometry ($\lambda/\Delta\lambda \sim 3$) in three spectral bands centred on approximately 250, 350 and 500µm. The three-bolometer arrays modules contain 43 (500µm), 88 (350µm) and 139 (250µm) hexagonally close-packed feedhorn-coupled NTD-detectors [5]. Each pixel has thus a conical feedhorn connected to a circular waveguide. A "spider-web" architecture [8], which consists of a mesh of silicon nitride (see Fig. 2), is then placed at the output of the feedhorn. It absorbs light and conducts the energy to the tiny thermistor that sits at the center of the web. The Neutron-Transmutation-Doped (NTD) germanium bolometer (indium bump bonded to the absorbers) then measures the local temperature [9].

Both bolometers are mounted thermally isolated from their own surrounding structure (~2K) and at an operating temperature of 0.3K provided by a dedicated closed cycle 3He sorption cooler.

### III. SPACE ENVIRONMENT AT L2

#### A. Expected before launch

The environment around L2 is relatively benign compared to those of lower orbits. Indeed, the L2 point is at such distance (1.5 millions km) from the Earth that the effect of geomagnetically trapped particles can be considered negligible. The only radiation sources that affect L2 are cosmic rays originating outside the solar system and high concentrations of particles emitted during solar events.

The rate of primary cosmic rays passing through a single PACS photometer array (16x16 pixels for a volume of $1.2 \times 1.2 \times 0.04$ cm$^3$) was estimated before the launch at 3 particles/s. Considering that primary particles passing through the surrounding materials also produce about 50-80% of additional events, either by nuclear reactions or gamma-ray emissions, the total expected rate was roughly 5 particles/s on each individual PACS photometer bolometer array [1].

#### B. Measured by SREM

SREM is a particle detector developed with the main purpose of permanent monitoring of the space radiation environment and providing alerts of radiation related hazards to the spacecraft and its payload. The reader is referred to the *Paul Scherrer Institut* (PSI) for more details about SREM [10]. Two models of SREMs are currently flying at L2 onboard Herschel and Planck spacecrafts, launched by ESA in May 2009. The SREM channel that is the most representative of cosmic protons is TC2, which records protons above 39MeV. The TC3 channel is also interesting but it is sensitive to lower energetic particles (protons > 10 MeV and electrons > 0.5 MeV), and thus more sensitive to solar events. The TC1 channel is sensitive to protons > 20 MeV and electrons > 2 MeV and looks very similar to TC3. The evolution of TC2 and TC3 raw count rates is displayed on Fig. 3 for the period between the launch and June 2011.

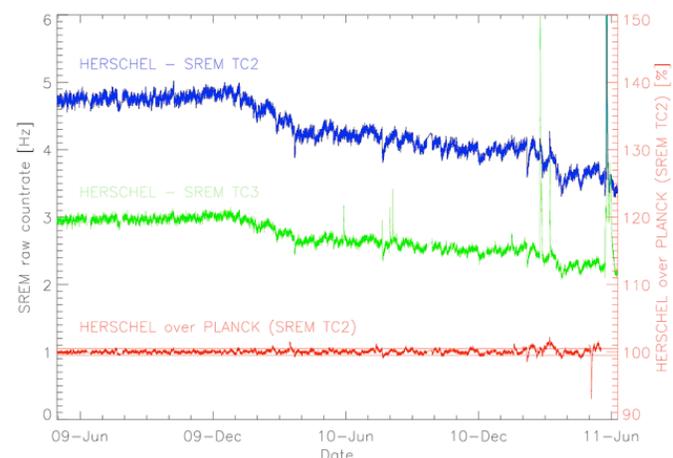

Fig. 3. The blue and green solid curves (units on the left vertical axis) correspond respectively to the raw countrates of TC2 channel (protons > 39 MeV) and TC3 channel (protons > 10 MeV, plus electrons > 0.5 MeV) recorded by SREM onboard Herschel since the launch date (mid of May 2009). The red curve (units on the right vertical axis) corresponds to the ratio of the count rates (TC2 channel) recorded by SREM onboard Herschel over those recorded by SREM onboard Planck during the same period. The two horizontal red lines delimit a variation of ±5% on both sides of a value of 1. The spikes show few solar flares.

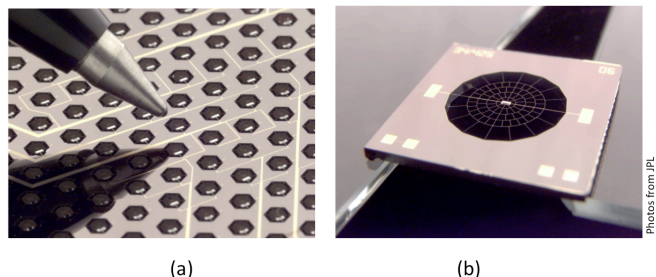

Fig. 2. (a) View of the SPIRE bolometer array with a pen for scale (pixel size 725µm). The bolometers are biased and read out via the gold leads on the silicon wafer. (b) A zoom-in on a single silicon nitride micromesh ("spider web") bolometer. The spider web mesh absorbs submillimeter radiation and germanium thermistor in the center detects the radiation.
(Pictures from http://casa.colorado.edu/~jglenn/research/spire.html)

By showing the evolution of the ratio of count rates measured onboard the Herschel over those measured onboard Planck, Fig. 3 also illustrates that both SREM measurements are very similar in spite of being measured by two independent particle monitors. This reinforces the confidence we can have in the measurements provided by SREM.

Since both spacecrafts were launched slightly before the last solar minimum (around January 2009), the cosmic ray flux was expected to decrease as soon as solar activity would start again. This is visible on Fig. 4 showing the ambient proton rate measured by SREM that started to decrease roughly at the beginning of 2010, i.e. roughly one year after the last solar minimum.

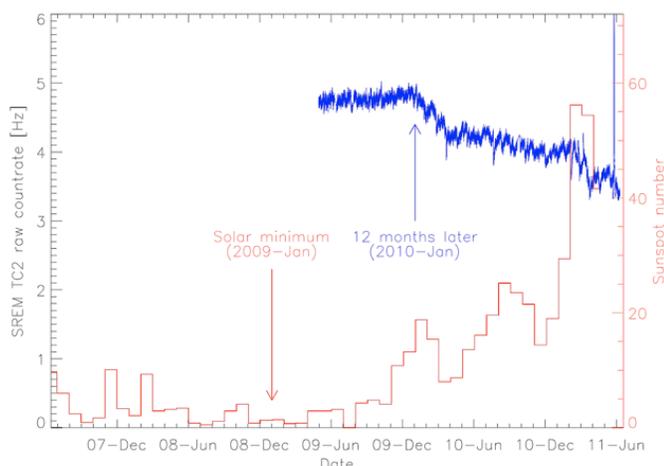

Fig. 4. Anti-correlation between the proton ambient rate at L2 and the solar activity: the blue curve represents the raw value of TC2 channel (protons above 39 MeV) recorded by SREM since the launch of Herschel (mid of May 2009), while the red curve represents the evolution of solar activity (number of sunspots), showing that the rate of cosmic particles started to decrease roughly 12 months after the solar minimum.

## IV. THE GLITCHES OF THE PACS Si-BOLOMETER

While traversing the bolometer, the high-energy particles ionize the atoms or molecules, which they encounter along their tracks. The thermalisation of this ionisation in the bolometer produces then a signal transient with time duration and amplitude dependent on the Linear Energy Transfer (LET) of the incident particle. The glitches detected by PACS bolometer can be divided into three families depending on their temporal profile and the tracks' positions in the sensitive part of the bolometer [1]:

- Type-A: *the positive glitches (~87%)*. They appear when the particle hits the grid of the bolometer. This is rapidly thermalized by phonons running in the entire structure and the local temperature increases (e.g. see Fig. 5-A). The amplitude is proportional to the LET corresponding to the deposited energy by the particle in the bolometer. The higher the LET, the higher the amplitude.
- Type-B: *the negative glitches (~13%)*. The negative glitches occur when the particle hits the inter-pixel wall where the common heat sink and the reference thermometers are located. The thermal disturbance would unbalance the bolometric bridge and cause the signal to drop. Depending on the absorbed energy they can appear in multiple pixels through a thermal cross talk (e.g. see Fig. 5-B and Fig. 5-C).
- Type-C: "*fader*". The faders generally deal with variation of the base line with a recovery time quite long (few minutes). These rare events appear when the particle hits the readout circuit, certainly because of the small cross-section of the transistors. (This effect is not investigated in this paper.)

The upper plot of Fig. 5 shows the time sequences of an entire multiplexing line corresponding to 16 pixels. The glitches are clearly identified as positives spikes (e.g. Fig. 5-A) or more rarely negative spikes (e.g. Fig. 5-B). We observe that the bolometer's properties (response, noise or gain) are not affected by numerous glitches. We do not have electrical cross talk.

The *positive glitches* have a rapid rise time (inferior to ~50ms equivalent to 2 frames with a 40Hz readout sample) and the decay time varies with the amplitude and the time constant of the bolometer (~24ms [1]) for a duration of few tens of ms (>150ms in Fig. 5-A).

Although the cross-section of the wall is rather important (due to a larger volume of a factor of ~1000) the temperature elevation, when the particle hits the inter-pixel wall, is smaller due to a larger thermal capacitance in comparison with the grid thermal properties. The *negative glitches* have thus a short time duration and a relative limited amplitude. Also the negative glitches are the less frequent ones (~13%). However when the LET is sufficiently high the thermal perturbation spreads on contiguous pixels and kinds of spots are observed on the matrix. To some extent, the size of these spots is also proportional to the deposited energy in the inter-pixel wall. Fig. 5-B plots the timelines on the pixels affected by one particle (the hit is arbitrary put at time=0s) and Fig. 5-C shows the image (at t=0s) on the matrix of this thermal cross talk.

Depending on the LET, overshoots or undershoots are also observed for the strongest glitches. Since the electronic passive filters are sufficiently fast (frequency~1280Hz) in the multiplexing circuit we cannot expect slow perturbations from the electrical chain. The origin of the signal perturbation could be due to an unbalance of the thermo-electric bolometric bridge when the LET is important. The heat propagates in the grid or the walls through the beams (see Fig. 1) and could cause the increase or decrease of the signal for a few frames, before reaching the equilibrium state depending on the thermal time constants of the system.

Fig. 6 shows the distribution of the intensity of the glitches with the MMT deglitching approach (Multi-resolution Median Transform [11]). The analysis has been performed with the set of data extracted from the observation of Abel-2218 during the calibration phase. This observation was performed at a dedicated 40Hz sample rate with one bolometer array only from the blue channel. The observed distribution of glitches is a power law of the form *glitchNumber* = *glitchPeak*$^{-b}$ with b=-1.7. Note that this power law does not directly represent the energy distribution of the cosmic rays at L2, but rather their energy spectrum weighted by the response of the bolometer, which is not linear in the full dynamic range. Note that MMT algorithm is quite aggressive as it tries to dig faint glitches out of the instrumental noise and below the noise limit of 10 ADU. Taking this into account the integrated number glitch is 3.3 counts/s/array (volume ~ $1.2 \times 1.2 \times 0.04$ cm$^3$) and the full-integrated number is 4.5. This measure is consistent with the expected 5 particle hits per second per array derived in [1].

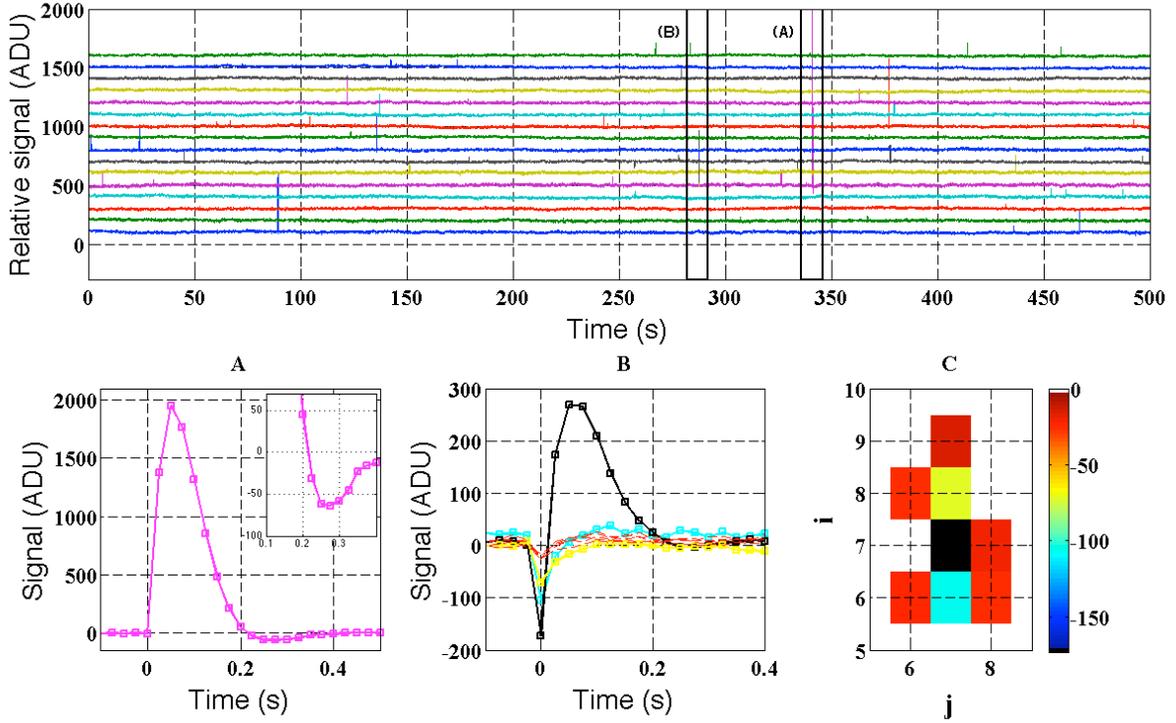

Fig. 5. The upper plot shows a PACS photometer time sequence (with a duration of 500s) of an entire multiplexing line corresponding to 16 pixels (40 Hz readout sample performed during calibration phase). We subtract the median on each time sequence and arbitrary add 100*i ADU on the signal (i from 1 to 16 corresponding to the pixel address). In figure A are shown the highest glitches of this example with a decay time (~150ms). The zoom shows the undershoot that follows the decay. The beginning of the glitch is arbitrary put at time=0s. In figure B are shown an example of thermal cross talk (box B in upper plot) where 8 pixels are then affected by the energetic cosmic ray hit in the inter-pixel wall. The time of the hit is arbitrary put to 0s. Figure C then illustrates the image (at t=0s) on the matrix (zoom on contiguous pixels) of this thermal cross talk as a spot. Only the thermal perturbations are plotted, the rest of the signal (noise) is put to zeros. i and j are the pixel address. The black pixel shows the strongest glitch.

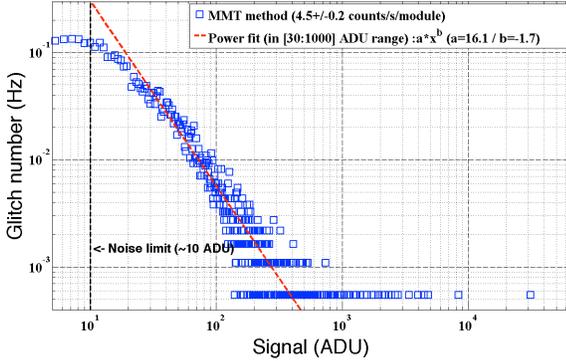

Fig. 6 Glitch number per second per PACS bolometer array as a function of the glitch intensity (maximum value of the positive or negative glitches).

## V. THE GLITCHES OF THE SPIRE FEEDHORN-COUPLED NTD-GERMANIUM BOLOMETER

The glitches of SPIRE bolometers are similar to the PACS ones. Sudden signal pulses are observed when cosmic-ray particle or photon are absorbed causing a rapid rise in temperature. The decay time depends on the thermal time constant determined by the thermal and electrical characteristics of the bolometer. The bolometer's properties (responsivity and noise) are unaffected after the thermal perturbation.

Two types of glitches are observed in the SPIRE detector timelines:
- Type-Single: these *single* events are observed in individual bolometer at a given time.
- Type-Concurrent: the *concurrent* glitches are seen simultaneously in all of the bolometer's thermistors, resistors and darks pixels of the array at a given time. It is due to ionizing hit on the silicon substrate that supports all the detectors in a given array (frame hits) [5].

*Single* glitches, are less common than *Concurrent*, however, they tend to be much stronger (up to thousands sigma). Even though easier to detect, determining the relaxation time is not straightforward and as result, unmasked glitch tales create obvious artefacts in SPIRE photometer maps. Currently, the automated SPIRE photometer pipelines apply a conservative, but inefficient 8-sample mask, following a *Single* glitch,

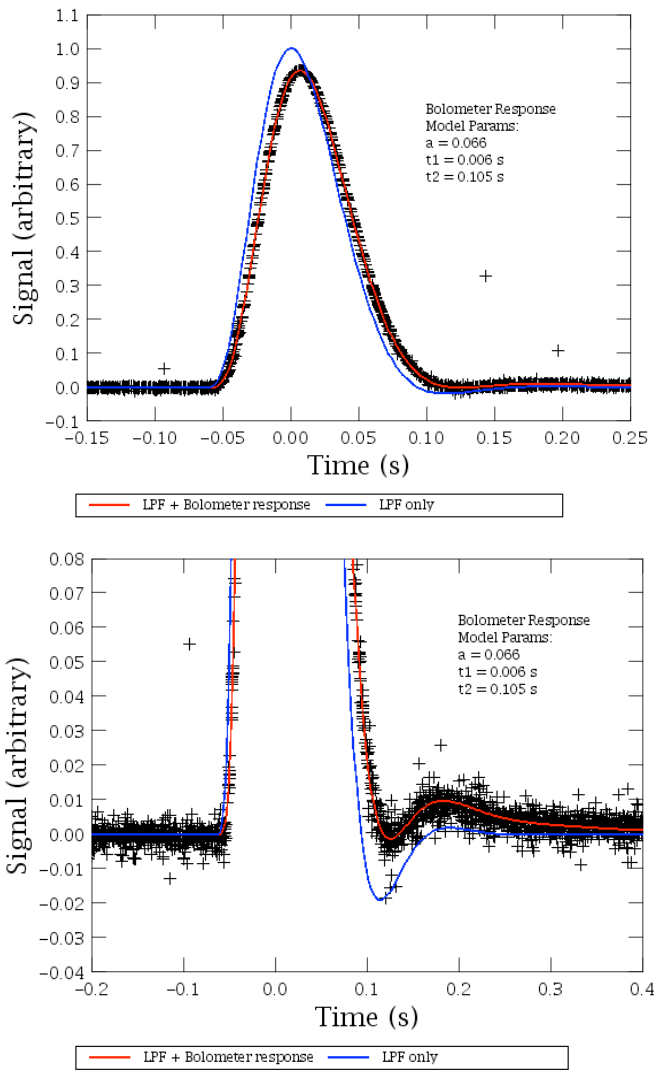

Fig. 7. Upper figure: The average response of a photometer bolometer (SPIRE PSWB3 detector) to an ionising radiation particle derived from the co-addition of many glitches (crosses) compared to the modelled impulse response of the electronics and bolometer with (red) and without (blue) a slow time response component. Lower figure: A zoom to an undershoot with the modelled response of the electronics.

however work is taking place to create a more appropriate adaptive masking procedure.

The *Single* glitches can be described very well with a bolometer transfer function that includes a second order slow time constant of approximately 0.1s. It has not been possible so far to identify whether this slow-time constant is optical or not (Fig. 7). This model also shows that undershoots observed for the highest glitches (or high LET) may come from the electrical active filters located in the electrical chain of the bolometer.

*Concurrent* glitches are comparatively more frequent than *Single* glitches, however they rarely are stronger than four sigma of the signal timeline. For this reasons such glitches have relatively smaller (nevertheless measurable) effect in maps. Typical rates of Concurrent glitches for PSW (250μm), PMW (350μm) and PLW (500μm) arrays are ~0.7, ~0.17 and ~0.5Hz respectively, for SPIRE only scan mode (18.6Hz sampling). For SPIRE/PACS parallel mode, these values are ~0.04, ~0.1 and ~0.35Hz. Unlike *Single* glitches, *Concurrent* glitches are better described with a bolometer transfer function that dos not include a slow time constant.

## VI. GLITCH RATE ON DETECTORS ONBOARD HERSCHEL

### A. Glitch rate on PACS Ge:Ga photoconductors

Compared to the bolometers, the significantly larger (1.5x1.5x1.0 mm$^3$ germanium crystal elements) individual detector pixels of the PACS spectrometer Ge:Ga detector array are susceptible to accordingly higher glitch rates within the same cosmic radiation environment. While the SREM instrument is located outside the Herschel cryostat, on the service module panel, pointing away from the sun, the detectors in the instruments are provided with additional shielding by the cryostat wall, its internal shields and instrument specific material geometries in the respective detector blocks. Reference [2] conducted the preparatory ground radiation tests of the stressed PACS Ge:Ga detectors under representative shielding conditions with respect to the in-flight Herschel case. Incoming 70 MeV protons are attenuated under these conditions down to energy of 17 MeV arriving at the detector crystal. Therefore we expect that only protons above ~50MeV would dominate the glitch rates on the PACS spectrometer detectors. The most comparable counter in the SREM instrument is therefore the TC2 channel, which is sensitive to all proton and heavy ion energies above >39 MeV [12]. The other two main channels and respective sub-channels TC1 and TC3 are sensitive as well to lower energetic particles and partly also to electrons. A comparison of radiation counts with these SREM detectors is therefore not straightforward and we therefore expect to see the best correlation with the TC2 channel.

Fig. 8 shows the Herschel SREM TC3 and TC2 channels throughout the mission since the launch on 14 May 2009 until July 2011. For each day on which PACS spectroscopy observations have been carried out, a daily mean glitch rate is derived from a single detector pixel in blue spectrometer channel, which is shown in the same plot. In order to detect a single glitch, a second order polynomial is fitted to the signal integration ramp (1/8th of a second, sampled at 256Hz) of the cold readout electronics ([2] and references therein). This fit is subtracted from the integration ramp and the standard deviation is calculated for the result. As soon as this standard deviation is larger than a defined threshold, a glitch is counted. The threshold is chosen sufficiently large to prevent false counts from electronic noise and other ramp discontinuities, like intended signal transitions by the chop-nod observing strategies.

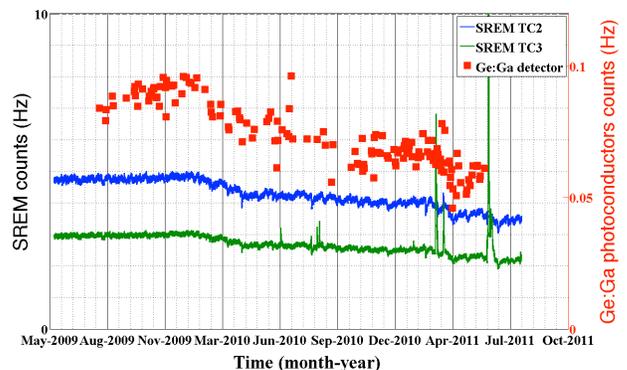

Fig. 8. The Ge:Ga PACS photoconductor glitch rate Vs. SREM TC2 and TC3 channels. The SREM TC2 and TC3 channels are smoothed to about 1 hour bins. The red squares are the daily mean glitch rate on a blue Ge:Ga detector whenever PACS spectroscopy was operated. Note that single outliers may be due to poor statistics (only few observations) or non-standard detector settings for calibration or engineering purposes during those particular operational days.

The overall trend in glitch rate (Fig. 8) on the blue PACS detector correlates well with both SREM channels (See Fig. 12 and Fig. 13), however only TC3 is sensitive to particle energies below 39 MeV. The few solar flare events detected during this period of time cause substantial variations on the TC3 rates, but only little increase in TC2 counts. This means that the energy spectra of the particles from those flares reaching Herschel were dominated by energies lower than 39 MeV.

In order to verify the conclusion of ground based irradiation tests of the PACS photoconductors, namely the rate is mainly dominated by particles with higher energies, two solar flare events have been analysed in more detail during PACS spectrometer operation. The two largest solar flares within the Herschel mission occurred on 8/9-Mar-2011 and 7/8-Jun-2011 and by chance, during significant parts of the flares, PACS spectroscopy observations had been carried out. Fig. 9 shows the PACS blue Ge:Ga glitch counts averaged to bin sizes of 10 minutes together with the SREM TC2 and TC3 channels. While the TC3 channel did show a substantial increase in counts during the March event, the TC2 channel counts remained almost stable. In contrast, during the June solar event both SREM channel counts increased, the TC3 channel by a factor of about 6 and the TC2 channel by about a factor of 3. At the same time the measured glitch rate on the Ge:Ga detectors increased as well by a factor of 3, thus confirming the expectation that the germanium crystals inside the PACS spectrometer are affected only by particles with energies at least above 39MeV.

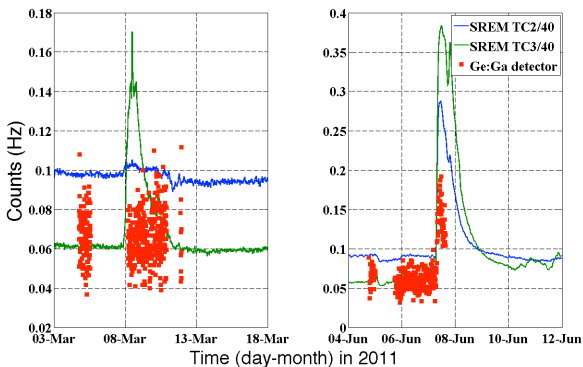

Fig. 9. Two solar flare observed with the Ge:Ga PACS photoconductor. The blue and green lines are the count rates by SREM channels TC2 and TC3 (divided by a factor 40 and smoothed to bins of 10 minutes) during the two largest solar events. The red squares are the count rates on blue PACS Ge:Ga detector binned to 10 minutes time resolution. The count rate is correlated with the TC2 channel (blue line).

### B. Glitch rate on SPIRE feedhorn-coupled bolometers arrays

The *wavelet deglitcher* [13] was used to identify glitches on all observations performed in *POF5* (Photometer Large Map) observing mode executed by SPIRE from August 2009 until June 2011. All observations were done with the same bias voltage. The glitch rate is then measured on each scan line (observation mode).

Fig. 10 shows the glitch rate (in Hz) on the 250µm bolometers array. We measure a big variation from one scan to the other, which means the current deglitcher detects false glitches. When the scan observation mode is operated the glitches are more difficult to distinguish from sources and false detections are certainly among them.

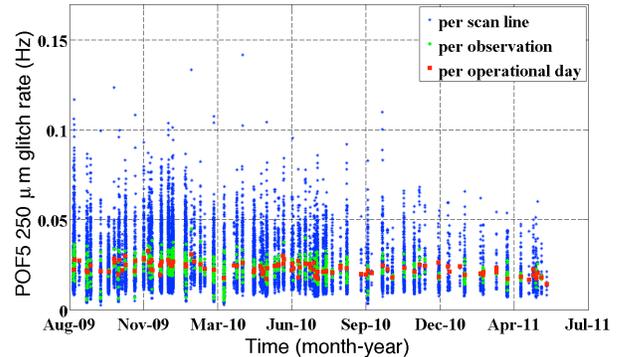

Fig. 10. The glitch rate on SPIRE feedhorn-coupled NTD germanium bolometers array (Photometer Large Map POF5 250µm bandwidth)
The blue spots are the count rates measured per scan line. The green circles are the average counts per observation and the red square the average counts per operational day.

Fig. 11 shows the average glitch rate (per operating day) for the SPIRE bolometer arrays (in the three-band) together with the SREM TC2 and TC3 channels. Despite the big variation measured from one scan to the other we observe a slight decrease correlating with the TC2 channel and corresponding to the increase of the solar activity (see Fig. 4).

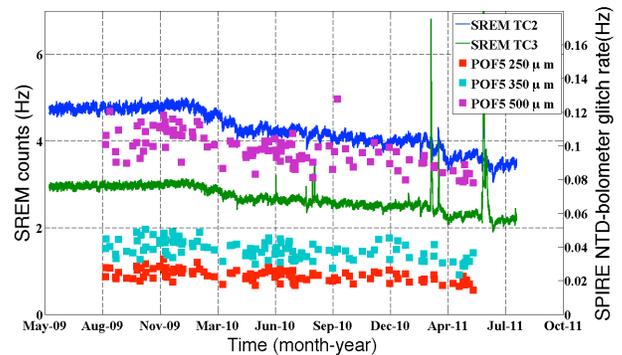

Fig. 11 The average glitch rate (per operating day) for the SPIRE feedhorn-couple NTD-bolometer arrays (Photometer Large Map POF5 in the three-band). The SREM TC2 and TC3 channels are smoothed to about 1 hour bins. The red, cyan and violet squares are the glitch rates per operational day on the SPIRE bolometers (250µm, 350µm and 500 µm bandwidth respectively).

### C. Glitch on PACS Si-bolometers arrays

To measure systematically the glitch rate on PACS photometer observations, we cannot use the wavelet method as this leads to false detection when the spacecraft scans on bright compact sources and ridges. We instead use a technique called second-level deglitching. It consists in projecting all the detector readouts on the sky and then, for each sky pixel, identifying the outliers in the stack of readouts that project onto it. This has been demonstrated as the most robust technique to flag out glitches while preserving actual sky signal. There however are a few caveats. First it is best to project the detector frame on a grid of sky pixels, and thus the stack we filter for outliers is not a stack of detector pixels but a stack of projected pixels. When an outlier is found, we flag in detector space all the pixels that contribute to this sky pixels, thus we likely overestimate the number of detector pixels hit by a particle. Then the detection of outliers can only be done once the low-frequency noise has been removed, which we do by applying a median-based high-pass filter to the data with a sliding window of 500 readouts, and our threshold to detect glitches is a function of the noise in the data. Should this noise level vary systematically, our ability to detect glitches would suffer from systematic bias. There is fortunately no indication that this is the case.

We have then sought the database of PACS observations for all scan maps posterior to posterior to Observation Day 156 (on 17 October 2009), which is the start of the routine phase observations, so that the instrument set-up is common to all observations. Detection with the second-level deglitching method is obviously easier when the same area of the sky is observed many times. Therefore we build our list of observations by first discarding observations shorter than 500s, and longer than 10000s. The lower limit is to reject very small observations where the coverage factor will be small, while the upper limit is to stay with observations that are reasonable to process, as second-level deglitching is obviously a time-consuming process. In the remaining observations we selected those that either had a repetition factor larger than 3 (i.e. the whole scan is repeated at least 3 times) or a distance between scan leg lesser than 50", so as to ensure a large number of individual readouts per sky pixel. This results in a list of approximately 3500 observations as of June 2011.

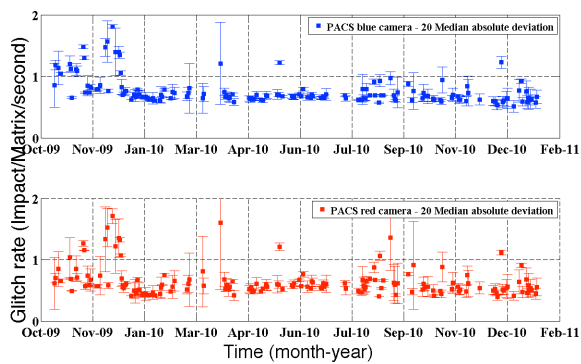

Fig. 12 Glitch rate on the two cameras of the PACS photometer. An impact is all the pixels detected, which is associated with a edge or an angle in both spatial directions and in temporal direction.

Each of the observation is then scanned for glitches and all the outliers pixels are flagged. Since a single impact by a particle can affect more than one detector pixel, and can last more than one readout, and since the second-level deglitching method can flag as outliers pixels that only happen to be next to a pixel hit by a particle, simply counting the number of flagged detector pixels is not correct to go back to the particle hit rate. We thus apply a grouping algorithm that identifies actual impact as group of neighbouring pixels in the cube of flagged pixels. Neighbourhood is allowed through the edges as well as through the summit of the pixels, as well as through time (i.e. successive readout). Therefore we end up with the actual number of impacts registered by the bolometers during the observation. We then divide this number by the duration of the observation, and the number of matrices in the focal plane (8 for the blue side, and 2 for the red side) in order to arrive at a number of impacts per second and per matrix (See Fig. 12). Unlike the PACS spectrometer we do not observe a decrease of the count rate. Although we measure big variation on few observations the averaged glitch rate is relatively flat except for the beginning of the mission and we cannot currently observe any correlation with the SREM channels (e.g. TC2) with the PACS Si-bolometer.

We can also measure the impacts of the glitches on the observation time for the two focal planes. Fig. 13 shows that the data loss is inferior to ~0.2% for the PACS blue and red cameras.

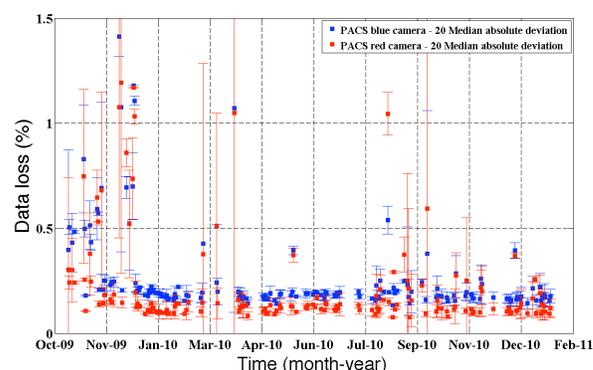

Fig. 13 Data loss in percentage on PACS photometer. The impacts of the glitches on the PACS photometer (2048 pixels for the blue camera and 512 pixels for the red one - 40 Hz onboard readout with 4 successive images averaged) is inferior to ~0.2% (or 4 pixels/image for the blue camera).

## VII. Correlation With SREM Data

Fig. 14 shows the particle counts (arbitrary unit) for the SPIRE bolometers and the PACS Ge:Ga photoconductors during the same epoch. On the right side the plots illustrates the cross-correlation functions with the SREM TC2 values.

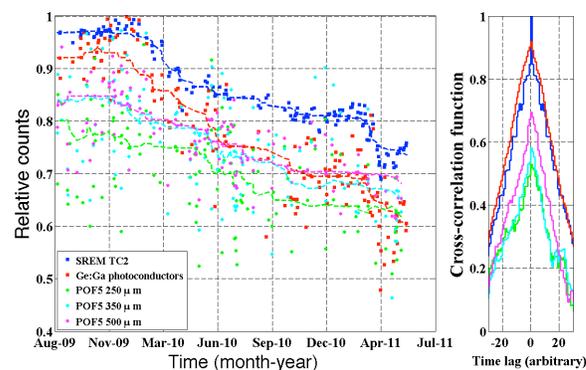

Fig. 14 The normalized glitch rate (average per operating day) for the SREM TC2, the SPIRE (in the three-band) and PACS bolometers, and the PACS Ge:Ga photoconductors. We plot only the values where PACS or SPIRE (or both) were operated. The dashed lines are the smoothed sample data with a mean-based filter with a sliding window of 20 operating days. On the right side the plots shows the cross-correlation functions (XCF) between the bolometers channels and the SREM TC2 channel. For information the blue line shows the auto-correlation of the SREM TC2 values (XCF=1 at time lag =0).

Although the overall trend seem to show a decrease, we can currently only suggest a correlation with the particle count measured by the TC2 channel. Only the glitch detected by the PACS spectrometer is very well correlated. Fig. 15 shows the glitch rate measured on detectors as a function of the TC2 channel. We operated a linear regression of each sample data showing the relative confidence with the correlation coefficient R. More operating days will be useful to complete the analysis and to fit to the SREM channels (e.g. TC2) and the decrease of the glitch rate. A complete analysis shall be possible at the end of the mission foreseen in December 2012.

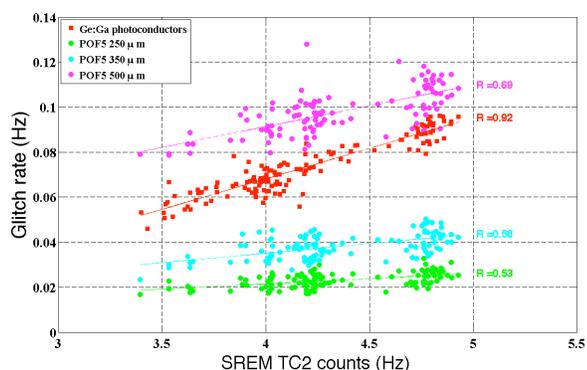

Fig. 15 The Glitch rates of the bolometers channels Vs. the SREM TC2 channel. We plot the glitch rate (average per operating day) for the SREM TC2 and the SPIRE feedhorn-couple NTD-bolometer arrays (Photometer Large Map POF5 in the three-band) and the Ge:Ga photoconductors. The dashed lines are the linear regression (order 1) of the sample data.

## VIII. SUMMARY

The space environment around the Sun-Earth Lagrangian point L2 was described with the Standard Radiation Monitor onboard the Herschel Space Observatory since the launch in May 2009. We showed that the count rate is similar between Herschel and Planck spacecrafts both around L2. The count rate measured by SREM monitors decreases from ~5 to ~4 Hz (TC2 channel which is sensitive to all proton and heavy ion with energies above >39 MeV), which may be caused by the increase of the solar activity.

We defined the different glitches observed on the two types of far-infrared bolometers inside SPIRE and PACS photometer. Some sudden signal pulses were observed with some characteristics depending on the thermo-electrical proprieties of the bolometer and also depending on the tracks' position in the sensitive part of the bolometer. We also observe some « undershoots » or « overshoots » of the signal for the strongest glitches in the two types of bolometers. While the origin may come from electrical active filters for the SPIRE feedhorn-coupled NTD-germanium bolometers, these signal variations is certainly due to an unbalance of the thermo-electric bolometric bridge with the PACS Si-bolometers. In the future we could try to understand this variation with a more realistic Si-bolometer model [14].

We also monitor the glitch rate of the SPIRE feedhorn-coupled NTD-germanium bolometers, the PACS Si-bolometers and the PACS Ge:Ga photoconductors from May 2009 to June 2011. The corresponding decrease was observed only on the SPIRE and PACS spectrometer detectors. The glitch rate measured on the PACS Si-bolometers is relatively flat during this epoch. While we don't observe any variation during the solar flare in March 2010, we observe a variation of the particle rate with the PACS Ge:Ga photoconductor for the solar flare in June 2010. That shows the glitch rate on the Ge:Ga crystals inside the PACS spectrometer are correlated with the TC2 channel, sensitive to particles of energies above 39 MeV.

We also measured the impacts on the observation time of the PACS photometer. The data loss is inferior to ~0.2% whereas it is inferior to 1% for the SPIRE photometer [4].

In this paper we have gathered different information from the detectors onboard Herschel. We need more operating days for a better statistics in the analysis, but we can already conclude that the radiation environment at L2 has no severe impact on the observation time at least with those far-infrared bolometers.